\documentclass[conference]{IEEEtran}
\usepackage{cite}
\usepackage{amsmath,amssymb,amsfonts}
\usepackage{algorithm,algorithmic}
\usepackage{graphicx} 
\usepackage{textcomp}
\usepackage{xcolor}
\usepackage{subfig}

\usepackage{makecell}

\def\BibTeX{{\rm B\kern-.05em{\sc i\kern-.025em b}\kern-.08em
    T\kern-.1667em\lower.7ex\hbox{E}\kern-.125emX}}

\begin{document}
\bstctlcite{IEEEexample:BSTcontrol}

\newcommand{\blue}[1]{\textcolor{teal}{#1}}
\newcommand{\green}[1]{\textcolor{violet}{#1}}
\newcommand{\purple}[1]{\textcolor{purple}{#1}}
\newcommand{\red}[1]{\textcolor{red}{#1}}

\title{
  Stream Compression of DLMS Smart Meter Readings\vspace{-0.5cm}
}
\author{
  \IEEEauthorblockN{Marcell Feh\'{e}r$^{1}$, Daniel E. Lucani$^{1}$, Morten Tranberg Hansen$^{2}$, Flemming Enevold Vester$^{2}$}
  \IEEEauthorblockA{
    $^{1}$Agile Cloud Lab, Department of Electrical and Computer Engineering, DIGIT, Aarhus University, Aarhus, Denmark\\
    $^{2}$ Kamstrup A/S, Skanderborg, Denmark\\
    \{sw0rdf1sh,daniel.lucani\}@ece.au.dk, \{mtr,flev\}@kamstrup.com 
  }
}

\maketitle

\begin{abstract}
  Smart electricity meters typically upload readings a few times a day. Utility providers aim to increase the upload frequency in order to access consumption information in near real time, but the legacy compressors fail to provide sufficient savings on the low-bandwidth, high-cost data connection. We propose a new compression method and data format for DLMS smart meter readings, which is significantly better with frequent uploads and enable reporting every reading in near real time with the same or lower data sizes than the currently available compressors in the DLMS protocol.
\end{abstract}


\section{Introduction}

In the last decade most electricity providers replaced their old meters (that had to be read by a technician) with smart connected devices, which are configured remotely and upload readings at predefined intervals without human assistance. Smart meters have numerous advantages beyond saving costs of manual readings: high granularity load monitoring, better anomaly and fraud detection, preventive maintenance or the ability to reduce billing frequency, just to name a few. 

The configuration and data upload formats for electricity smart meter communications are focused on flexibility, resulting in highly configurable but verbose protocols, collectively known as \textit{Device Language Message Specification} (DLMS\footnote{https://www.dlms.com/}). Data compression options have been part of this de-facto standard for a long time, but the supported algorithms no longer provide adequate size reductions considering recent service demands. As utility providers want to receive consumption readings more frequently (e.g. every 15-60 minutes instead of once a day), the basic assumptions of the built-in DLMS compressors about payload size do not hold any more, and compression rates decline significantly \cite{feher2020smartmeter}. When a single reading is uploaded at a time, the existing compressors do not compress at all. 

In this paper we continue our efforts to find better data compression methods for DLMS-based smart meters, leveraging a technique called \textit{Generalized Deduplication}\cite{vestergaard2019bounds}. We propose a new algorithm and companion data format that achieves the same level of compression when readings are uploaded in real time as the commonly used DLMS Null compressor uploading readings once a day. We evaluate our method on real life data captured over a 9 months from 95 households, provided by a major danish smart meter manufacturer, Kamstrup A/S.

The paper is organized as follows. After reviewing existing literature about smart meter reading compression in Section \ref{sec:relatedwork}, we describe the important properties of the DLMS data format and built-in compressors, as well as the concept of generalized deduplication in Section \ref{sec:background}. Next, we introduce our proposed compression scheme in Section \ref{sec:contribution}, and evaluate it in Section \ref{sec:evaluation}. In Section \ref{sec:conclusions} we summarize our findings.

\section{Related Work}\label{sec:relatedwork}

Compressing smart meter data is a relatively new field, as the need for more frequent uploads is a recent trend. In \cite{feher2020smartmeter} we introduced generalized deduplication-based compressors using a single GD transformation function and configuration, and processing the encoded APDU bytes sequentially. We now extend this method with positioning GD chunks to the underlying structure of the data, and add support for different settings in a single compressed packet.

Other methods in this field typically require a model of power consumption to compress. 
Lee \textit{et al.}~\cite{lee2019unified} proposes compressive sensing with empirical modeling of the electricity consumption of different devices in homes of residential customers.
In \cite{kraus2012optimal} the authors describe a method using the LZMA algorithm and predictive modeling, but their technique requires a dictionary and additional preprocessing before compression can start.
A different approach in \cite{abuadbba2017gaussian} aims to minimize the parameters required to represent meter readings. They use a method based on Gaussian approximation.
Our proposed compressor does not require any preprocessing or modeling step.
The authors of \cite{blalock2018sprintz} propose a forecasting-based stream compressor that encodes the difference between the prediction and the actual measurements.

\section{Background} \label{sec:background}

In order to describe the context of the proposed compression method, we start by a brief introduction of the DLMS data format of smart meter readings, as well as the built-in compressors that our results will be compared to. We also give a high level introduction to the underlying technique of our algorithms: generalized deduplication.

\begin{figure*}[tbp]
  \centerline{\includegraphics[width=\textwidth]{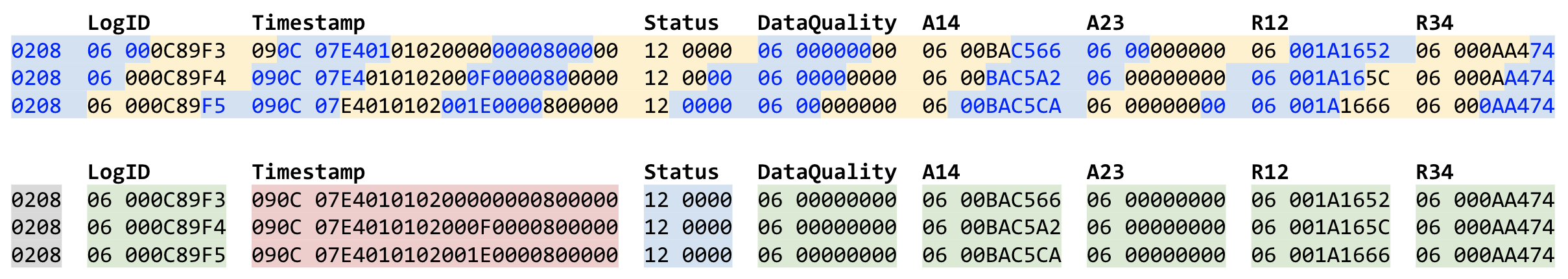}}
  \caption{ When GD chunks are not aligned with the similar regions of data, compression potential is poor(above). Pattern-based GD moves chunks to where similarity is expected (below). }
  \label{fig:chunk_alignment} 
  \vspace{-2mm}
\end{figure*}

\subsection{DLMS Data Format}

The DLMS protocol specifies the communication protocol between a smart meter backend and the meters both for configuration (server to meter) and reporting (meter to server). Meters typically record active and reactive energy consumption every 15 minutes, and upload them once or multiple times a day, depending on local regulations and utility provider preferences. 

Meters can be configured with multiple so-called \textit{Load Profiles} (LP), which determine what signals to record, how frequently, and when to upload the readings. Even though we only focus on compressing uploaded data packets from the Load Profile that records power consumption, there are other profiles as well, for example for anomaly detection or power quality measurements.

When a Load Profile trigger condition is met, the meter collects the corresponding readings since the last upload and constructs a data packet (called \textit{Application Data Unit} or APDU) by encoding the data values and adding a header that describes the LP. The header contents and internal structure can be heavily customized by the meter manufacturer and the utility provider, but the data buffer is only based on the included signals, therefore identical across providers. For this reason, our compression algorithm only processes the APDU data buffer, and assumes a different compressor handles the header.

\subsection{DLMS Compressors}

The DLMS protocol includes a few (optional) compression methods to decrease the APDU data buffer size, which we will compare our algorithms with.

\textbf{Null Data} is useful when repeating or strictly predictable values are present in the APDU. It works by replacing the actual value by a single NULL byte if it is identical to the previous one (for example the same reactive energy was measured twice in a row), or the current value was expected from the previous one (e.g., the timestamp was increased by the amount defined in the Load Profile). 

\textbf{Delta Array} aims to encode the difference of consecutive values on fewer bytes instead of the values themselves. It can efficiently compress the energy consumption columns, since it typically requires only 1 or 2 bytes to represent the delta, instead of 4 bytes for the original full values. Both of these methods use the previous reading as a basis for compression, therefore only effective when at least two readings are included in an APDU.

\textbf{V.44} is a general purpose statistical compressor that is included in the DLMS protocol suite. Unlike the previous methods, it processes the APDU as a continuous byte array, e.g. without any regards to the internal structure of the encoded readings.

\subsection{Generalized Deduplication}

Deduplication is a well known technique that decreases size by storing repeating data chunks only once. While it is great for data with identical segments (e.g file servers), it fails to compress even small changes between chunks. Vestergaard \textit{et al.} ~\cite{vestergaard2019generalized} extended this method with the possibility of deduplicating similar chunks. The resulting algorithm, called \textit{Generalized Deduplication} (GD) works by first transforming each data chunk with a user-supplied function to a pair of basis and deviation. Ideally, the basis stores the repeating information across chunks, while the difference is captured in the smaller deviation. As a result, similar data chunks produce the same basis, which can be deduplicated in the compressed representation. The decompressor applies the inverse transformation to basis-deviation pairs to reconstruct the original data. 

This compression method is highly efficient in scenarios where the data has high correlation and there exists a reversible transformation that can separate the variance from the constant part of each chunk. IoT data in general and smart meter readings specifically are very good fits for this technique. 
%
\begin{figure*}[tpb] 
  \centerline{\includegraphics[width=0.9\textwidth]{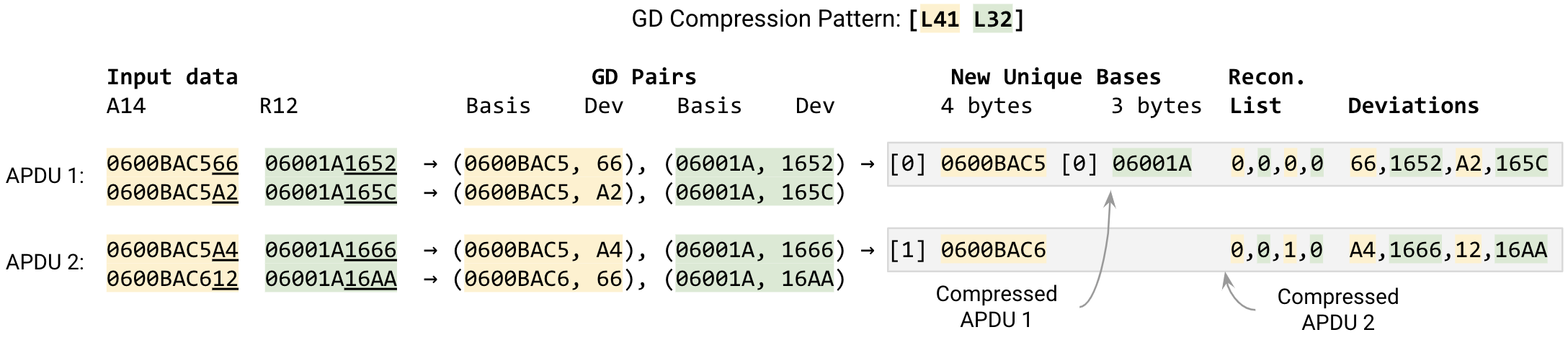}}
  \caption{ Data format of the pattern-based GD stream compressor. }
  \label{fig:data_format} 
  \vspace{-2mm}
\end{figure*} 

\section{GD Pattern-Based Stream Compression}\label{sec:contribution}

Due to the fact that the APDU header is customizable by the utility provider and differs significantly between companies, we focus our efforts to compressing only the Data sections, e.g. the encoded list of power consumption readings. Furthermore, since the header is constant for a Load Profile, it can be represented by a predefined ID or a hash digest, as we proposed in \cite{feher2020smartmeter}. Consequently, the best strategy is to split the APDU and compress the two sections with different approaches.

Similarly to general-purpose statistical compressors like V.44, in our previous works we used GD-based techniques on APDUs as continuous byte sequences, without leveraging their internal structure. This led to good compression rates and straightforward applicability over a wide range of APDU configurations. However, a lot of compression potential was left unexploited, as the deduplicated chunks were not aligned with the similar data sections in the APDU. Figure \ref{fig:chunk_alignment} illustrates this issue, as the 4-byte GD chunks are not aligned with the data columns. This skew does not eliminate compression completely, but GD basis duplications are rather accidental and rare than intentional and exploited every time.

Since the data section of an APDU follows a fixed, table-like structure where column widths are predictable\footnote{ The DLMS protocol guarantees that values are always encoded to a constant number of bytes, depending on the data type. }, it is possible to align the GD chunks to the similar byte regions (data columns).

\subsection{Pattern-based Generalized Deduplication}

To exploit this predictability, our new GD-based compressor allows the user to define a pattern of transformation function configurations that are applied to the input byte array (encoded APDU) continuously. This way it is possible to position the deduplicated chunks precisely to the data columns (see lower half of Figure \ref{fig:chunk_alignment})

The pattern is given as a compact string that consists of GD algorithms and their configurations. Each pair directly or implicitly defines the chunk size and the size of GD basis and deviation. For example, the pattern \textbf{\texttt{[H5 H6]}} prescribes using Hamming code with 5 parity bits, then Hamming code with 6 parity bits. This transformation function only has one parameter (parity bits), which determines the basis and deviation sizes (3+1 bytes for H5, 7+1 bytes for H6), therefore the chunk size as well (4 bytes for H5, 8 bytes for H6). During compression, the pattern is repeated until the end of the input data stream. 

To define the pattern used on Figure \ref{fig:chunk_alignment}, we cannot use Hamming code as the transformation function, since it does not have a setting for 3, 5 and 14 byte chunks. Furthermore, the highly variable bytes are always at constant positions in an APDU, something that Hamming code does not exploit very well, and a different transformation function is needed.

\begin{figure*}[t] 
  \centerline{\includegraphics[width=0.95\textwidth]{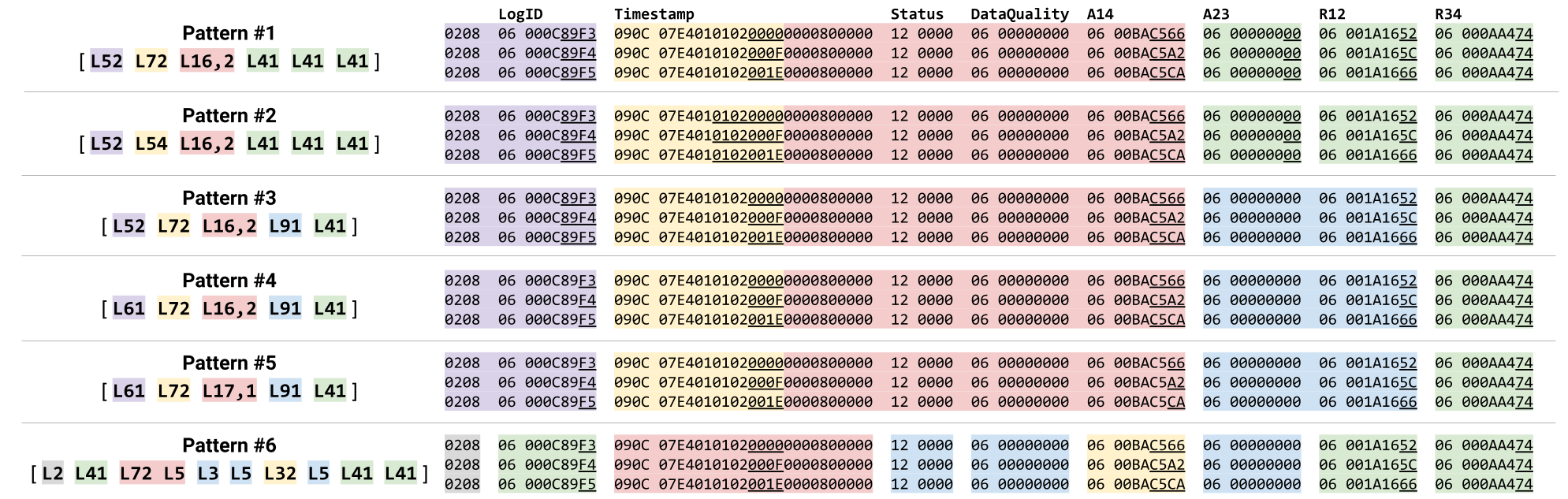}}
  \caption{ The evaluated GD patterns (deviation indicated by underline). }
  \label{fig:patterns_list} 
  \vspace{-2mm}
\end{figure*} 

\subsection{LastByte GD Transformation Function}

We propose a new transformation function for generalized deduplication that is applicable when the highly variable bytes of a chunk are at the end, and their length is known. 

\vspace{.5cm}
$\begin{array}{lcl}
    LB_{n, k}(chunk) & => & (chunk[0,n], chunk[n,k])\\
    LB^{-1}(base, dev) & => & base+dev
\end{array}$
\vspace{2mm}

During compression, the LastByte function  takes an $n+k$ byte chunk and splits it: the first $n$ bytes become the basis and the subsequent $k$ bytes form the deviation. The inverse function is used in the decompressor, which glues the basis and deviation back together to reconstruct the original data chunk. This simple transformation very efficiently produces identical bases and small deviations when applied to unsigned integers in big endian encoding, when the range of values, and therefore the number of highly varying bytes can be estimated well. All of these conditions are true for DLMS APDUs, and the computational simplicity also makes this function ideal for low powered embedded systems like a smart meter.

A LastByte-transformed chunk is encoded in a pattern string as $Ln,k$, or $Lnk$ when both parameters are single digits. A possible pattern for the coloring of Figure \ref{fig:chunk_alignment} is the following \textit{(note that the colored regions do not indicate where the basis ends and the deviation starts)}: 

\vspace{1mm}
\textbf{\texttt{[L20 L41 L14,0 L21 L32 L41 L50 L41 L41]}}
\vspace{1mm}

Pattern-based generalized deduplication with the LastByte transformation function on APDUs has numerous advantages over our previously proposed compressors, which used Hamming code and a single chunk size:
\begin{itemize}
  \item Deduplicated data chunks are positioned exactly to the data columns, not skewed any more.
  \item LastByte enables extracting the bytes with high variance as deviation, instead of the bits dictated by Hamming code.
  \item The encoded readings can be covered without any gaps, therefore no data padding is necessary.
\end{itemize}

In Section \ref{sec:evaluation} we evaluate different patterns to see how to best separate the encoded readings to chunks, and where to position the deviations to maximize compression gain.

\subsection{Serialization Format}

The compression process and data format of the compressed data format is illustrated on Figure \ref{fig:data_format}, where for two APDUs, both containing two readings of only the A14 and R12 signals are processed. First, DLMS-encoded input data is converted to a list of basis-deviation pairs according to the pattern. Next, new unique bases are added to in-memory arrays. These arrays form the compressor state that needs to be stored between transmissions. To transmit the new APDU, only the new GD bases and the (basis index,deviation) pairs are necessary. Since the decompressor also knows the pattern (it can either be sent in the first APDU, or inferred from the Load Profile), both the index and deviation lists can be transmitted without any separators. For example, if the \textbf{\texttt{[L41 L32]}} pattern is used, the decompressor picks bases from the 4-byte and the 3-byte arrays in an alternating manner. Furthermore, since both the compressor and decompressor knows how many unique bases are in each array, the number of bits to address them can be minimized, and packed tightly into bytes. In our example, the whole Base Reconstruction List (that holds indexes to the unique base arrays) fits into 4 bits per APDU. Finally, the list of deviations are also serialized without delimiters, because the decompressor knows from the pattern how many bytes to read at a time. 
This data format heavily leverages the information present in the pattern itself, and achieves very low overhead for serializing the new information of the next compressed APDU.

\section{Performance Evaluation}\label{sec:evaluation}

We measure the compression of the pattern-based generalized deduplication method using a real life dataset provided by Kamstrup A/S, a major smart meter manufacturer in Denmark. It includes readings data from 95 consumers over 9 months between 2020 January and September. Every 15 minutes the smart meters recorded a reading with 4 quantities, listed in Table \ref{table:apdu_contents}. A single reading is encoded to a total of 49 bytes, including DLMS type codes that precede each value and additional overheads.

\begin{table}[htbp]
  \caption{Contents of a reading}
  \centering
  \begin{tabular}{c l l} 
   \textbf{Order} & \textbf{Name}  & \textbf{Data type} \\
   1 &  Log Id     & 4 byte unsigned  \\ 
   2 &  Timestamp & 12 bytes \\
   3 &  Log Status & 2 byte bitmap \\
   4 &  Data Quality & 4 byte bitmap \\
   5 &  Active energy consumption (A14)  & 4 byte unsigned \\
   6 &  Active energy generation (A23)  & 4 byte unsigned \\
   7 &  \makecell[l]{Reactive energy consumption\\phases 1 and 2 (R12)}  & 4 byte unsigned \\
   8 &  \makecell[l]{Reactive energy consumption\\phases 3 and 4 (R34)} & 4 byte unsigned \\
  \end{tabular}
  \label{table:apdu_contents}
\end{table}

As the motivation of our research is the desire of utility providers to receive consumption data more frequently, we focus the compressor evaluations to smaller APDUs. Namely, we will calculate compression gains at the sizes listed in Table \ref{table:apdu_sizes_horizontal}. Typical data upload periods have been decreasing from 24-12 hours to 2-6 hours. The ultimate goal is to lower this to 1 hour and under, the most valuable being real time upload of every single reading. 

\begin{table}[htbp]
  \caption{Evaluated APDU sizes}
  \centering
  \begin{tabular}{r c c c c c c c } 
    \textbf{Readings/APDU:} & 1 & 2 & 4 & 12 & 24 & 48 & 96  \vspace{1mm} \\
    \textbf{Upload period:} & 15min & 30min & 1h & 3h & 6h & 12h & 24h \\
  \end{tabular}
\label{table:apdu_sizes_horizontal}
\end{table}

To evaluate our compressor against the built-in methods and compare different patterns, we calculate the compression gain over the whole duration of the dataset, separately for each consumer, and report the average across all households. For the GD pattern-based stream method, we report the average after the first transient APDU. The compression gain is expressed as $1 - compression\ rate$ in percentage, where higher values are better (0\%: no compression, 100\%: data is reduced to zero bytes). We have replaced the V.44 compressor, which is patented and no implementation is freely available, with the LZMA algorithm. It is a fundamentally very similar statistical compressor, built into Python.

\begin{figure*}[htpb]  
  \centerline{\includegraphics[width=0.9\textwidth]{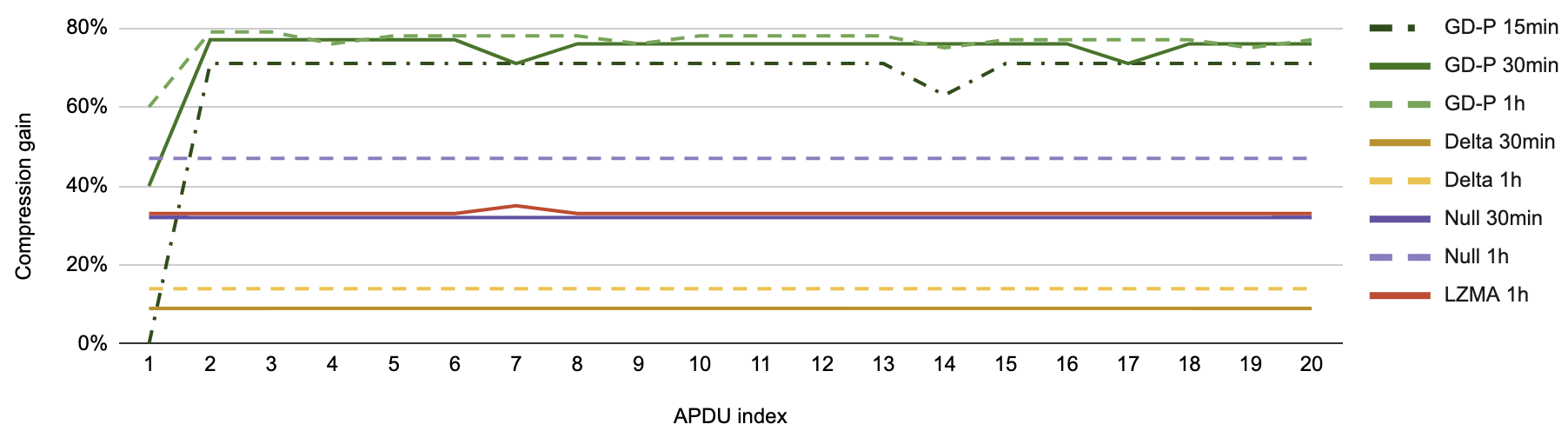}}
  \caption{ The GD-based stream compressor achieves its eventual compression rate already at the second APDU. }
  \label{fig:stream_convergence}  
\end{figure*}

We will evaluate the GD pattern-based compressor in Stream Mode (as described in \cite{feher2020smartmeter}), where consecutive APDUs of a single meter are considered a data stream. With this technique the unique GD bases that repeat across APDUs are sent only once, the first time they appear. Data savings are realized two ways: when there are similar values within a data packet, and across packets from the same meter. Figure \ref{fig:stream_convergence} shows how the compression rate changes with successive APDUs. Our pattern-based stream compressor achieves very high compression already from the second APDU, even if every reading is uploaded immediately. The downside of this approach is the introduction of state in the compressor and decompressor that needs to be stored between data uploads.

\newcommand\Tstrut{\rule{0pt}{2.6ex}}         
\newcommand\Bstrut{\rule[-0.9ex]{0pt}{0pt}}   

\subsection{Different Patterns}


Since the pattern-based compressor is highly flexible, we can easily evaluate variations that correspond to different assumptions about the dataset, and see which compresses better.
In the following, we describe five handcrafted and a generic pattern, and evaluate their performace. Figure \ref{fig:patterns_list} shows the pattern strings and illustrates how they map to a few encoded readings.

\textbf{Pattern \#1} exploits the constant \texttt{0208} type code (meaning: structure of 8 elements) in the beginning of each encoded reading, and merges it with the LogID in the first chunk. The last two LogID bytes form the deviation, which allows 256 consecutive readings to produce the same GD basis. The DLMS standard encodes the timestamp in a way that the highly variable hour and minute bytes are at offsets 8 and 9, so we define a 7+2 LastByte that covers the first half of the timestamp and captures these bytes as deviation. The rest of timestamp, together with the Status and Data Quality bitmaps are expected to be constant throughout the dataset. For this reason, the following chunk goes all the way to the end of A14, where the next highly varying bytes are expected. In this pattern we assume higher range of active energy readings and allow two bytes for the deviation. The last three data columns are processed as separate 5-byte chunks, where the one byte is reserved for deviation. This equals to the assumption of lower variance in the produced active energy and reactive energy quantities. \textbf{Pattern \#2} only differs in second rule: it assigns two additional bytes of the timestamp to the deviation, the ones that encode the day of the reading. \textbf{Pattern \#3} goes back to having only the hours and minutes as deviation, and assumes A23 to be constant across readings by merging it into the basis of the next chunk. Since very few households are feeding electricity back to the power grid, this change is likely to increase compression. \textbf{Pattern \#4} combines shorter LogID deviation, assumption of constant A23 and higher A14 variance, while \textbf{Pattern \#5} shortens the A14 deviation to a single byte. \textbf{Pattern \#6} is fundamentally different from the previous ones. It explores whether a fixed GD pattern can be assigned to each quantity, and locality is not exploited within the encoded reading. 
 
\begin{figure}[htpb] 
  \centerline{\includegraphics[width=0.95\columnwidth]{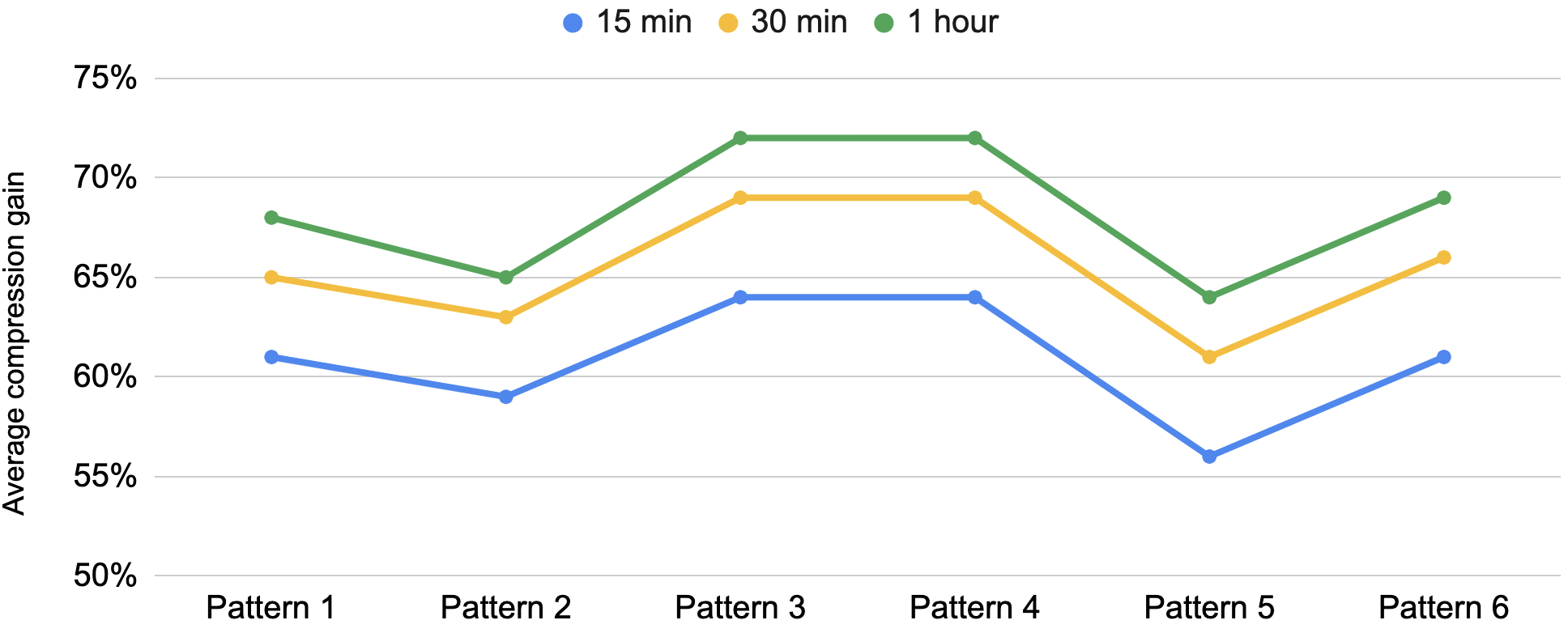}}
  \caption{ Average compression rates of the patterns, over the whole time period }
  \label{fig:patterns_results} 
\end{figure}

\begin{figure*}[htpb] 
  \centerline{\includegraphics[width=0.95\textwidth]{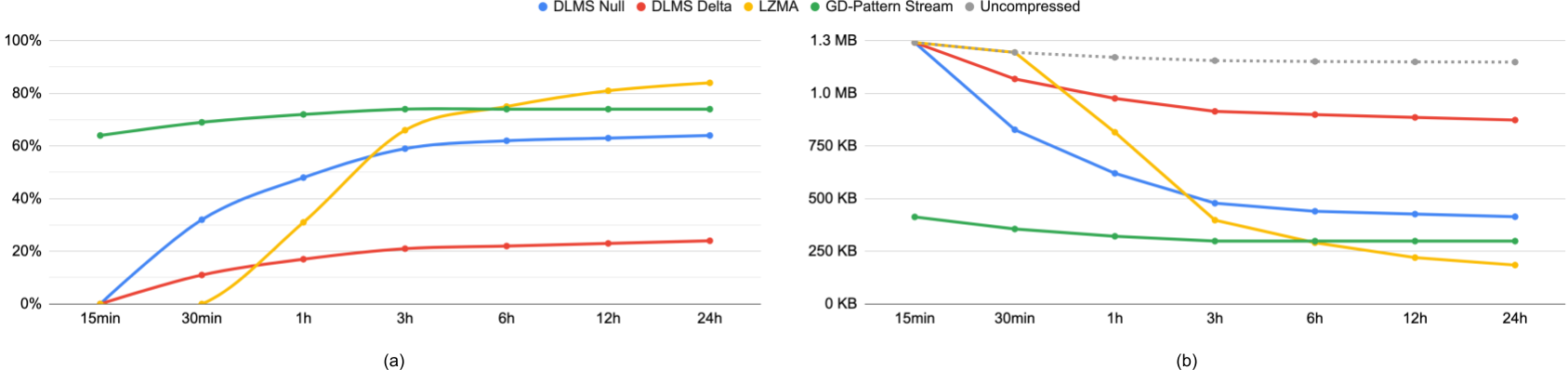}}
  \caption{ Compression gains (a) and total uploaded data (b) of the evaluated compressors. }
  \label{fig:compression_rates_sizes_all} 
  
\end{figure*}

The results show that some of our assumptions yield slightly better compression than others. Since patterns \#3 and \#4 are the best, the better choice about timestamp is to maximize the deduplicated basis and keep only the hours and minutes in the deviation. Also, it is best to separate only the last byte of the LogID, but A14 varies so much that two bytes deviation is the better choice. Since pattern \#6 achieves only 3\% lower compression rates than the best hand-optimized ones, it is safe to declare that automatic generation of patterns is a viable option. This proves the real life applicability of this compression method, since the GD pattern can be generated from the Load Profile definition, and does not require a specialist to construct the optimal pattern string on a case-by-case basis.

\begin{figure}[htpb] 
  \centerline{\includegraphics[width=0.95\columnwidth]{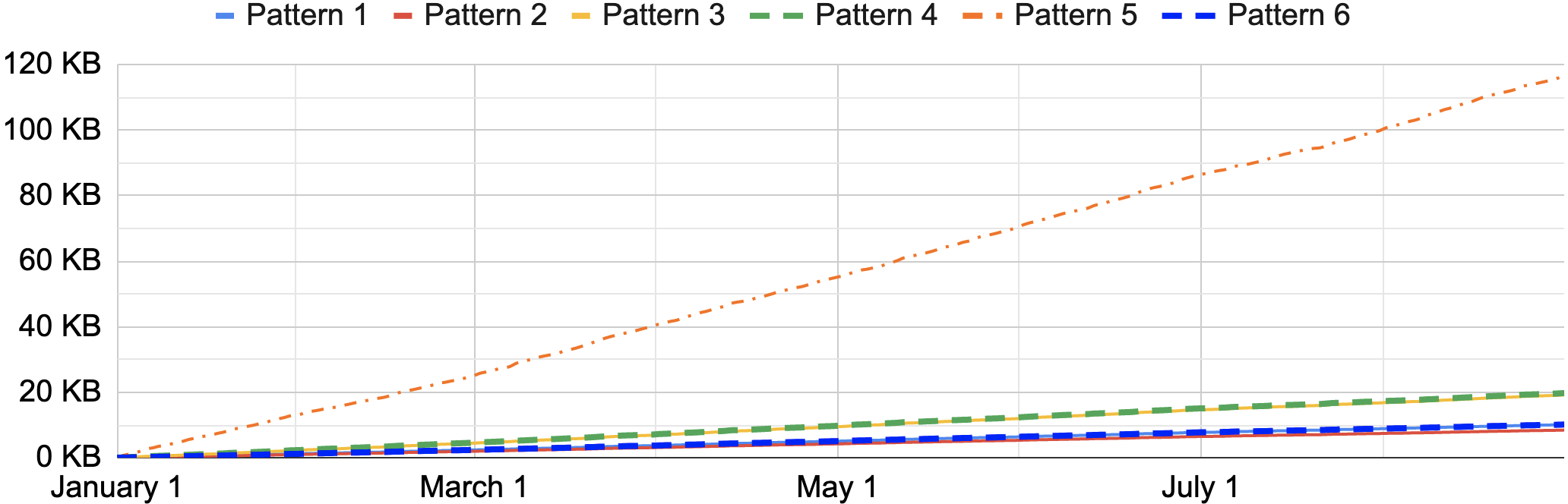}}
  \caption{ Size of the compressor state over time, per meter. }
  \label{fig:compressor_state_size} 
\end{figure}

Since the stream compressor achieves low transmission size by introducing a state of GD bases sent in previous APDUs, it is important to see how large this state is for each pattern. Figure \ref{fig:compressor_state_size} shows how the state grows over the 9 months time period. Note, that the state size is independent of the upload period, since the same amount of information is being transferred over time, the same GD bases form the state regardless of how frequently a new APDU is uploaded. The relative growth rate is constant with each pattern, and the memory requirements correspond to the compression gains: patterns with higher compression require larger state. 

Note, that our proposed compressor keeps old GD bases in the unique basis arrays indefinitely, therefore the memory requirement constantly increases over time. An interesting future research direction is to remove the unused bases from the state over time, either by reference counting or simply defining a sliding window and evicting old bases automatically. Both improvements can be implemented easily with the existing compressor architecture, and will be explored later.

\subsection{Comparison with DLMS Compressors}

The existing DLMS compressors fall short when APDU size is drastically reduced. Null and Delta methods use the first reading as the reference, and only process the consecutive ones. Therefore, their performance with only two and four readings per APDU is not satisfactory, and they do not compress when a single reading is uploaded.
Another issue with the built-in compressors, including V.44, is they only compress one APDU at a time, and not leveraging information from the previous data packets. 

The left side of Figure \ref{fig:compression_rates_sizes_all} shows the compression gains of the compressors over all 95 meters and 9 months (using pattern \#4 in the GD-Pattern method). The DLMS methods and general purpose statistical algorithms achieve satisfactory compression rates when the upload period is multiple hours, since they need a larger data buffer to be efficient. The statistics-based LZMA method is noticably good at 12 and 24 hour reporting periods, but due to its computational complexity, it is rarely used in real life. None of the methods currently available in the DLMS standard provide compression of a single reading. Our proposed compressor is able to reduce APDU size to 64\% when every reading is uploaded separately, and up to 74\% with larger upload periods.

In right side of Figure \ref{fig:compression_rates_sizes_all} shows the total amount of data uploaded by one meter over 9 months, and added the total uncompressed APDU data size as a reference. We report the complete APDU size (header and data together), where the header is represented by a 4-byte ID. Our GD-P method can process the header together with the APDU data by simply extending the pattern to the following: \textbf{\texttt{L40 [ L52 L72 L16,2 L91 L41 ]}}. The first 4 bytes (header ID) are treated as a GD basis and referenced/deduplicated together with all other 4-byte bases. Then, the rest of the pattern is repeated over the encoded readings, just like before. As a result, the header occupies 4 bytes in the first GD-P compressed APDU, and only a few bits in all subsequent ones, depending on the number of unique 4-byte bases. In the other compressors the header ID is repeated in each APDU.

The data shows that our proposed compressor is able to report each reading immediately with similar amount of data transfer (413 kB) than the best DLMS method with only one upload per day (DLMS Null, 414 kB), or a heavy statistical compressor reporting every 3 hours (LZMA, 398 kB). This compression efficiency allows utility providers to access near real time information with no additional data transfer cost.

\section{Conclusions} \label{sec:conclusions}
In this paper we have presented a new GD transformation function and a flexible, pattern-based extension of a previously introduced strem compression technique for smart meter readings. We have shown that the pattern can be automatically inferred from the Load Profile definition, making the method easily applicable at scale. The proposed compressor is highly efficient in reducing transfer sizes when data is uploaded frequently, and allows utility providers to upload every reading immediately.


\ifCLASSOPTIONcaptionsoff
  \newpage
\fi
\bibliographystyle{IEEEtran}
\bibliography{ref}

\end{document}